# Dense-Sparse Deep Convolutional Neural Networks Training for Image Denoising

Basit O. Alawode

*Electrical and Computer Engineering Department, Khalifa University,*
*Abu Dhabi, 127788, United Arab Emirates*
*100060517@ku.ac.ae*

Mudassir Masood*

*Electrical Engineering Department and the Center for Communication Systems and Sensing, King Fahd University of Petroleum and Minerals, Dhahran, 31261, Saudi Arabia.*
*mudassir@kfupm.edu.sa*

Recently, deep learning methods such as the convolutional neural networks have gained prominence in the area of image denoising. This is owing to their proven ability to surpass state-of-the-art classical image denoising algorithms such as block-matching and 3D filtering algorithm. Deep denoising convolutional neural networks use many feed-forward convolution layers with added regularization methods of batch normalization and residual learning to speed up training and improve denoising performance significantly. However, this comes at the expense of a huge number of trainable parameters. In this paper, we show that by employing an enhanced dense-sparse-dense network training procedure to the deep denoising convolutional neural networks, comparable denoising performance level can be achieved at a significantly reduced number of trainable parameters. We derive motivation from the fact that networks trained using the dense-sparse-dense approach have been shown to attain performance boost with reduced number of parameters. The proposed reduced deep denoising convolutional neural networks network is an efficient denoising model with significantly reduced parameters and comparable performance to the deep denoising convolutional neural networks. Additionally, denoising was achieved at significantly reduced processing time.

## 1. Introduction

Image denoising has been an active area of research for a long time as it is an integral step in many practical applications such as medicine, astronomy, etc. Noise is inevitable in images and videos captured by devices such as cameras, magnetic resonance systems, etc. Noise could be introduced due to some irregularities in the components that make up the system, the environment where the picture was captured (for instance, a foggy or rainy environment), etc. Therefore, the task of image denoising algorithms (denoisers) is to recover the clean image $X$ from a noise corrupted image observation $Y$. Usually,

$$Y = X + W \tag{1}$$

In (1), $W$ is referred to as the noise. The noise is usually independent of the image. To model this noise, random mutually independent entries drawn from the same probability distribution (i.i.d) is mostly used. The Gaussian (normal) distribution is a widely used distribution for modeling this noise. Therefore, $W$ is commonly assumed to be additive, white and Gaussian (AWGN) having a variance $\sigma^2$ and a mean equal to zero. Over the years, many denoisers have been developed and can be broadly categorized into two classes, i) Classical [Refs. 1, 3, 4, 6, 9, 11, 14, 17, 23] and ii) Neural Networks (NNs) based denoisers [Refs. 5, 16, 19, 25–28].

The classical denoisers take advantage of various mathematical concepts to perform denoising. These techniques can be non-patch [Ref. 4] or patch-based [Refs. 3, 6, 9, 14]. An image patch is a subset of an image. It contains several adjacent pixels cut out from the image to further enhance the image processing. Non-patch-based denoisers such as the non-local (NL) means algorithm [Ref. 4] operate directly on the image pixels. Particularly, the NL-means denoiser filters out noise from a target pixel by replacing it with the mean of all pixels in the image weighted by how similar they are to the target pixel. Patch-based denoisers use several patches, mostly overlapping, to perform denoising. A single image patch contains more information about an image than a single image pixel. This makes patch-based denoisers perform better than the non-patch-based counterparts. In many patch-based denoisers, patches are transformed from the spatial domain to a transform domain. A popular choice is to transform patches to a sparse domain. In this domain, denoising becomes a question of identifying the components which are due to noise and removing them. Curvelet [Ref. 21], wavelet [Ref. 7], contourlet [Ref. 8], discrete cosine transform (DCT) and discrete wavelet transform (DWT) [Ref. 18] dictionaries are some of the prespecified dictionaries used for patch transformation. Dictionaries can also be learned from the image patches themselves with the help of dictionary learning algorithms such as K-singular value decomposition (K-SVD) [Ref. 2], the method of optimal directions (MOD) [Ref. 10], etc. Given the dictionary, transformation is done using pursuit algorithms such as the orthogonal matching pursuit (OMP) [Ref. 2] and support agnostic Bayesian matching (SABMP) [Ref. 15] algorithms. Block-Matching and 3D filtering (BM3D) [Ref. 6], Collaborative Support-Agnostic Recovery (CSAR) [Ref. 3] and K-SVD [Ref. 9] denoisers have used patch transformation in the sparse domain to perform denoising.

The recent successes of deep learning (DL) algorithms, which are mostly implemented using NNs, is due to the availability of enormous data and processing power. These have seen the performance of the state-of-the-art (SOTA) classical denoisers surpassed by the DL-based counterparts and have inspired researchers to remodel the classical denoisers using NNs. Of such denoisers is the BM3D Network (BM3D-Net) [Ref. 24] which converts the classical BM3D denoising steps into a set of convolution layers. Another recently proposed denoiser is the Deep K-SVD [Ref. 20] which takes advantage of NNs to learn the values of some parameters in the classical K-SVD denoiser's pipeline. However, the performance of these denoisers is still not up to par with that of pure DL denoisers.

Several pure DL-based denoisers have been proposed [Refs. 5, 19, 25–28]. Of note is the DL-based denoiser - the deep denoising convolutional neural networks (DnCNN) proposed in [Ref. 26]. Due to its deep architecture and better performance, it has inspired several other denoisers such as the fast feed-forward network (FFDNet) [Ref. 27], and the enhanced deep convolutional neural network (EDCNN) [Ref. 28]. The DnCNN uses several layers of convolutions and takes advantage of the advances in DL regularization techniques such as batch normalization and residual learning to obtain an improved denoising performance surpassing many SOTA denoisers in performance. However, its deep architecture results in a huge set of trainable parameters. Do we really need such a deep network to achieve comparable performance?

In this paper, we show that by employing a different training strategy, we can significantly reduce the number of layers of the DnCNN and still achieve comparable denoising performance. The resultant architecture is termed reduced DnCNN (RDnCNN). Specifically, we employ the dense-sparse-dense (DSD) training approach [Ref. 10] to train the RDnCNN. This approach has been utilized for well-known architectures such as GoogLeNet, VGG-16, ResNet-16, DeepSpeech and achieved better performances compared to their normally trained counterparts. The RDnCNN was able to achieve comparable performance at a significantly reduced network parameters and denoising time compared to the full DnCNN. The rest of the paper is organized as follows. The DnCNN approach for image denoising is described in Sec. 2. We discuss the DSD training approach in Sec. 3. In Sec. 4, we present the proposed algorithm. Finally, the results and discussion are presented in Sec. 5.

## 2. DnCNN for Image Denoising

As DnCNN [Ref. 26] forms the base of our work, we provide the details of the DnCNN denoiser in this section. The basic architecture of the DnCNN is shown in Fig. 1. The network is composed of 17 convolution layers to perform image denoising. All layers except the output layer is activated with a Rectified Linear Unit (ReLU) activation. This deep structure is able to gradually separate the image from its noisy observation resulting in a significant advantage over other architectures. Batch normalization [Ref. 22] is applied to all hidden convolution layers to address the problem of internal covariate shift which ensued as a result of stacking several layers of convolution. As opposed to learning the clean image, the noise, termed the residual image, is learned by the architecture. This is then removed from the noisy image to get the denoised image. This concept follows residual learning strategy [Ref. 13] which, combined with batch normalization helped in boosting the model's performance. As aforementioned, the main disadvantage of the DnCNN is the enormous computational complexity that arise from the need to train an extensive set of parameters.

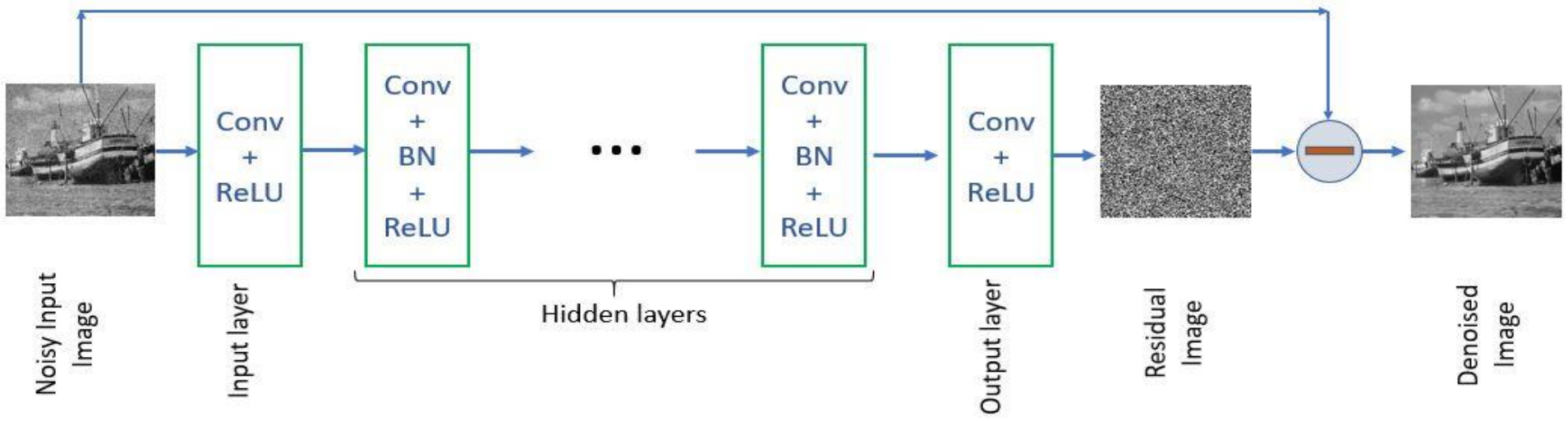

Fig. 1. The DnCNN Network Architecture.

## 3. DSD Neural Network Training

In this section, we provide the detail of the dense-sparse-dense (DSD) training approach. The authors in [Ref. 12] through the training of well-known NN architectures such as the GoogLeNet, VGG-16, ResNet-16, and DeepSpeech observed that some network weights actually tend to impact negatively on the performance of the network. They devised the DSD approach in which such weights were strategically masked to achieve appreciable performance boost in these networks. The DSD approach is a three-step process as described below.

**Step 1: Dense Training** - Given a normal (dense) NN architecture as shown in Fig. 2 (left), the network is trained normally with all its parameters available for training. As all weights are utilized for the training, it is termed dense training. During this training, the network is able to learn the set of weights necessary to achieve good performance. This is essentially the approach used in training many NN architectures.

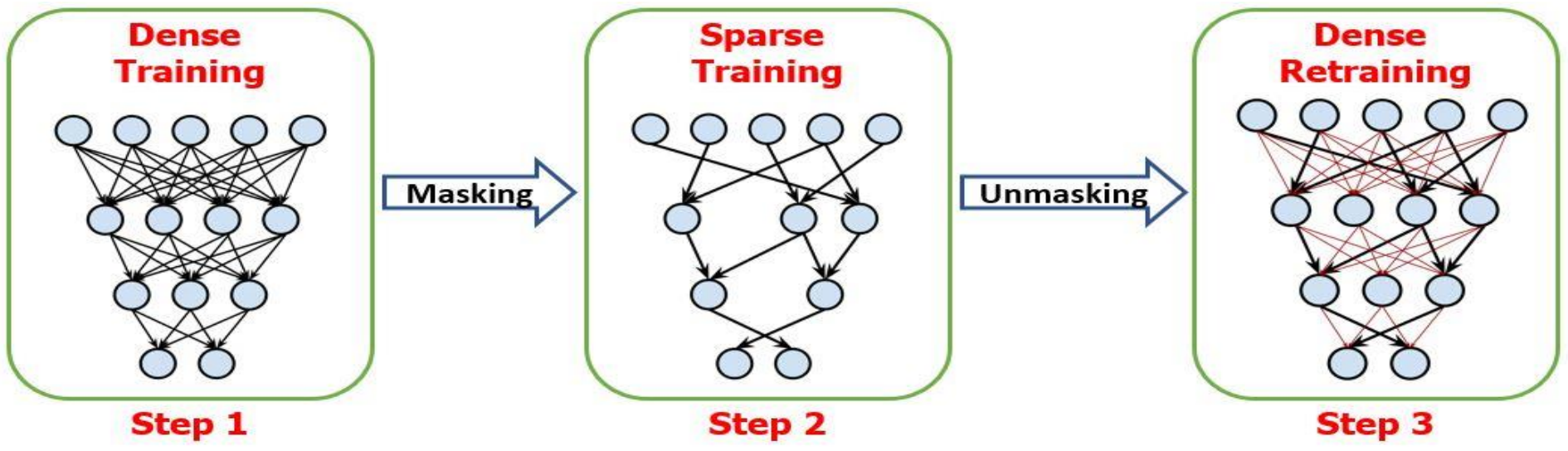

Fig. 2: The Dense-Sparse-Dense (DSD) Training Flow

**Step 2: Sparse Training** – This step is for fine-tuning the learned weights of the network in step 1 as some these learned weights could have negative impact on the performance of the network. To achieve this, the trained weights in each layer of the network are ranked in terms of their absolute values and a certain bottom percentile from them are masked, i.e., set equal to zero. This results in a network with sparse set of trainable weights (Fig. 2 - middle). The resulting network represents a sparse version of the original network. Training is then continued with this sparse network. It is to be noted that the masked weights do not partake in the training process. After training, the sparse network has better optimized set of weights and the performance of such network have been shown to be better than their dense counterparts in all tested network types. It should be noted that this step is different from dropout used to combat overfitting in NNs. The resulting sparse network in this training step is actually the network on which performance will be evaluated on during test time if this network happens to be the final network. This is opposed to dropout which releases the dropped weights during the test time.

**Step 3: Dense Retraining** – Lastly, the masked weights in step 2 are released and training is continued on the once again dense network (Fig. 2 – right). This is done to further perturb the already efficient sparse network in step 2, perhaps, a better global minimum could be achieved. As we have observed, not all networks yield appreciable performance improvement with this step. This is because the network obtained after the sparse training in step 2, for most networks, is already at the global optimization minimum. However, this can only be known after some ablation studies on such network.

## 4. Proposed RDnCNN

Encouraged by the training approach of the DSD, the importance of the DnCNN to denoising using NNs, and the need to have high performing shallower networks, we apply the DSD training strategy to a reduced DnCNN architecture termed the RDnCNN. The proposed RDnCNN architecture follows Fig. 1 and contains only 12 layers compared to DnCNN with 17. With the RDnCNN and DSD training approach, we show that comparable performance can still be achieved by using a shallower architecture. The RDnCNN culminates in a parameter reduction from 557,057 for the DnCNN to 371,777 for the RDnCNN.

The main performance defining strategy of the DSD is the sparse network training of step 2 above. At this stage, the network is able to rid itself of potential weights that could negatively impact its performance. Our ablation studies showed that in our application of the DSD to the RDnCNN, the dense retraining in step 3 only yielded a negligible improvement in performance. The increase in training time caused as a result of the dense retraining is not enough a price to pay for the negligible performance improvement. As such, our final RDnCNN model is obtained after the sparse training.

We first train the RDnCNN using the full weights in a dense approach and then mask the lower 15% of the ranked dense-trained weights. The resulting sparse model is further trained to obtain the final RDnCNN sparse denoiser. It should be noted that while the total number of weights in the RDnCNN network is 371,777, the effective number of weights taking part in the denoising is only 316,010 (15% less 371,777).

### 4.1 Choice of Number of Layers and Sparsity Rate

To decide the best tradeoff between the number of layers and sparsity rate, we train networks with different depths (number of layers) and sparsity rates and report the results in terms of the PSNR and SSIM in Table 1 below. This table shows the denoising performance on noisy images with $\sigma = 25$. We can observe from this table that the appropriate tradeoff is at number of layers of 12 and sparsity rate of 15%. Above these values, no appreciable performance improvement was observed. This has informed our choice of 12 layers and 15% sparsity rate.

To cater for the increased training time as a result of the DSD training approach, we reduce the number of training epoch from 50 utilized in the DnCNN to 40 for the RDnCNN. This ensures that we have a comparable training time. The training data and all the other parameters such as the learning rate are kept the same as utilized by the DnCNN.

Table 1. Choice of Layers and Sparsity Level

| | | | **Sparsity (Masking %)** | | |
|---|---|---|---|---|---|
| | | | **10%** | **15%** | **20%** |
| **Layers** | **10** | PSNR | 29.88 | 29.98 | 29.96 |
| | | SSIM | 0.88 | 0.89 | 0.89 |
| | **12** | PSNR | 29.98 | **30.18** | 30.16 |
| | | SSIM | 0.89 | **0.91** | 0.91 |
| | **15** | PSNR | 30.10 | 30.18 | 30.17 |
| | | SSIM | 0.91 | 0.91 | 0.91 |

## 5. Results and Discussion

Extensive experiments and comparison with other denoising algorithms were performed. Our choice of the DnCNN was due to its better performance compared to other SOTA denoisers [Ref. 2–5, 8]. As our algorithm is a direct modification of the DnCNN algorithm, we compare our performance with that of DnCNN only. We used Python-based NN Keras library running on top of TensorFlow library to develop both models. The base model for the DnCNN model can be downloaded from https://github.com/cszn/DnCNN. Training and testing were performed using 2 Nvidia GTX 1080 Ti GPUs running on a Linux-based cluster with an allocated 32GB of RAM. Training took about 230 minutes for the DnCNN and about 240 minutes for the RDnCNN. This is a comparable training time given the training approach employed for the RDnCNN.

### 5.1. *Comparison with DnCNN*

We present a summary of the comparison of the parameters of the RDnCNN and DnCNN in Table 2. These parameters were used to achieve comparable denoising performance. We can observe a reduced number of layers which results in a 16.8% (including masked weights) reduction in the number of trainable parameters. Due to the efficiency of the RDnCNN network, the average denoising time is significantly reduced by 57%. Additionally, the proposed RDnCNN has a significantly reduced average denoising time compared to the DnCNN.

Table 2. Parameters Summary

| | **DnCNN** | **Proposed RDnCNN** |
|---|---|---|
| Number of layers | 17 | 12 |
| Number of training epochs | 50 | 40 |
| Model Sparsity | - | 15% |
| Training time | 230 minutes | 240 minutes |
| Average Denoising Time | 35 seconds | 15 seconds |
| Trainable parameters | 447,057 | 371, 777 |

We tested both models on several images utilized in literature for denoising. To prepare the experiment images, we added white Gaussian noise of known variances to the clean images. For a fair comparison with the DnCNN, noise with σ = 15, 25, and 50 were considered. The noisy images were fed into both algorithms to obtain the denoised images. Denoised images were subsequently compared to the original image using commonly used image comparison metrics - the peak signal to noise ratio (PSNR) and the structural similarity index (SSIM). Results of some of these images are presented in Table 3. In the table, we show the PSNR and SSIM of the noisy image (before denoising), denoising with DnCNN, and with the proposed RDnCNN. To emphasize the effect of the sparse training, we also show the denoising results of dense only training to compare with that of dense-sparse (DS) training. We can observe that the performance of DS training surpasses that of the dense only training for all images. The proposed RDnCNN trained using the DS approach offers comparable performance to the DnCNN but with a noticeable reduction in the denoising time.

Table 3. Performance Comparison of Proposed RDnCNN with the DnCNN

| | | **Before Denoising** | **DnCNN** | **Proposed RDnCNN (Dense Only)** | **Proposed RDnCNN (Dense-Sparse)** |
|---|---|---|---|---|---|
| **Mandrill (σ= 15)** | PSNR (dB) | 24.60 | 32.35 | 30.07 | 32.34 |
| | SSIM | 0.72 | 0.93 | 0.90 | 0.93 |
| **Pepper (σ= 15)** | PSNR (dB) | 24.60 | 33.10 | 30.08 | 33.12 |
| | SSIM | 0.55 | 0.94 | 0.90 | 0.94 |
| **Boat (σ= 25)** | PSNR (dB) | 20.16 | 30.19 | 29.92 | 30.18 |
| | SSIM | 0.34 | 0.91 | 0.89 | 0.91 |
| **Cameraman (σ= 25)** | PSNR (dB) | 20.16 | 29.98 | 28.00 | 29.99 |
| | SSIM | 0.33 | 0.91 | 0.89 | 0.92 |
| **House** | PSNR (dB) | 14.14 | 28.00 | 27.71 | 28.00 |
| **(σ= 50)** | SSIM | 0.12 | 0.82 | 0.80 | 0.82 |
| **Barbara** | PSNR (dB) | 14.14 | 26.25 | 25.32 | 26.23 |
| **(σ= 50)** | SSIM | 0.20 | 0.80 | 0.79 | 0.80 |

Visually, as shown in Fig. 3, there are no observable degradation in the denoised images produced by the proposed RDnCNN when compared with that of the DnCNN.

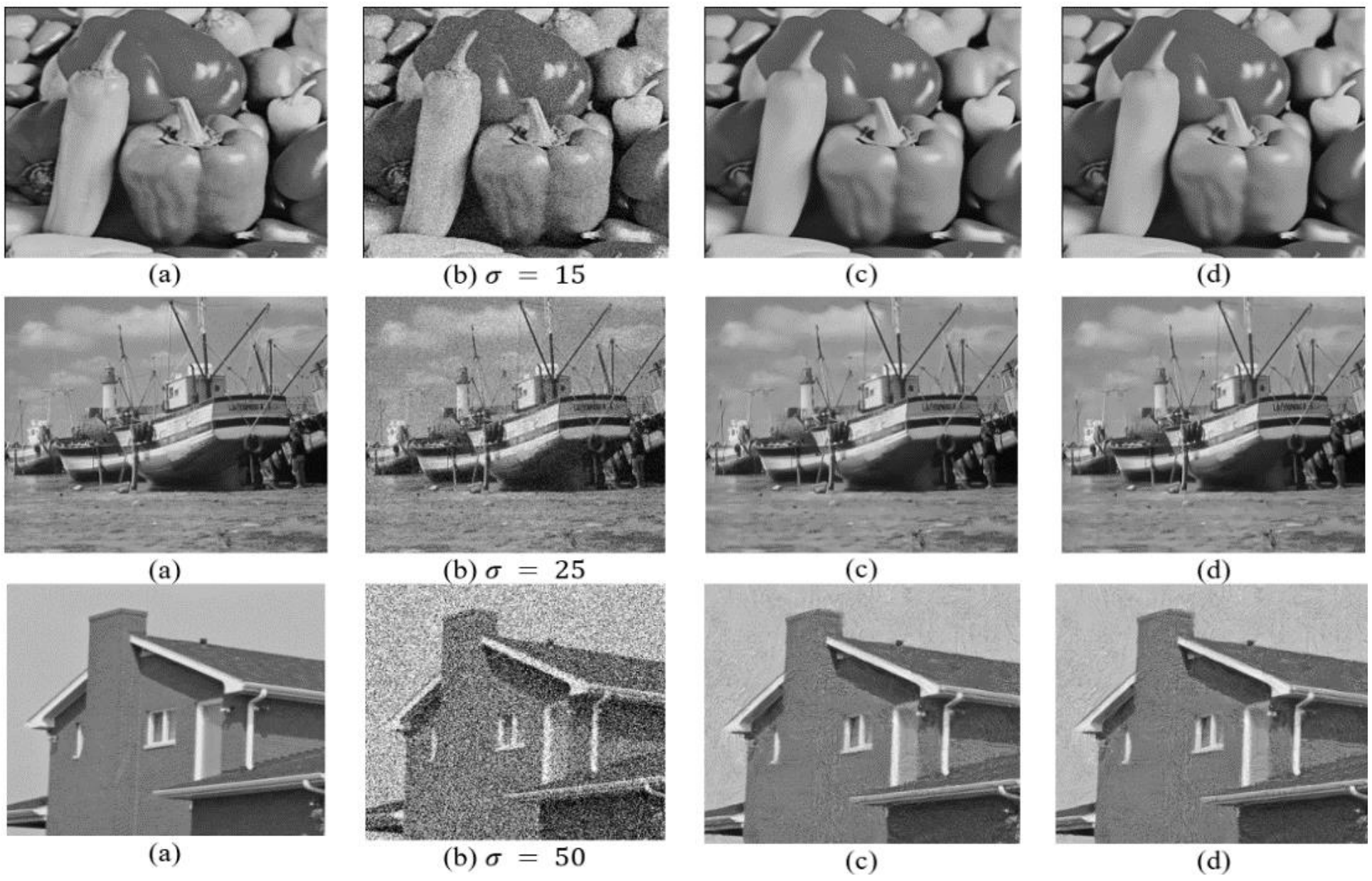

Fig. 3: Comparison of DnCNN with the proposed RDnCNN for different images at different noise levels. (a) Original images, (b) Noisy images, (c) DnCNN, and (d) Proposed RDnCNN.

## 6. Conclusion

In this paper, we successfully incorporate the DSD training approach into one of the SOTA DL-based denoising algorithms - the DnCNN. The choice of DnCNN was born out of the fact that its performance surpasses many SOTA denoising algorithms and has inspired several other DL-based denoising algorithms over the years. However, DnCNN achieves this performance at the expense of increased number of deep convolution layers resulting in a vast number of parameters to train. Our successful application of the DSD to the proposed RDnCNN resulted in a network with significantly reduced number of layers and parameters but with comparable denoising performance when compared with the DnCNN. Experimental results showed that the RDnCNN does not only perform comparatively close to the DnCNN, but also achieves faster denoising time. Further research will be done on the application of the DSD to other well-known neural network models in various applications and domains.

## 7. Acknowledgements

Part of this work was carried out by the first author (Basit Alawode) during his visit to King Abdullah University of Science and Technology, Thuwal, Saudi Arabia. The author expresses gratitude to the institution for their support and help with computational resources.

## References

1. M. Abedini, H. Haddad, M.F. Masouleh, A. Shahbahrami, Image Denoising Using Sparse Representation and Principal Component Analysis, *Int J Image Graph* (2022). doi: 10.1142/S0219467822500334.
2. M. Aharon, M. Elad, A. Bruckstein, K-SVD: An algorithm for designing overcomplete dictionaries for sparse representation. *IEEE Trans Signal Process* 54:4311–4322 (2006). doi: 10.1109/TSP.2006.881199.
3. M. Behzad, M. Masood, T. Ballal, M. Shadaydeh, T.Y. Al-Naffouri, Image denoising via collaborative support-agnostic recovery, *IEEE International Conference on Acoustics, Speech and Signal Processing*, pp 1343–1347, (2017).
4. A. Buades, B. Coll, J.M. Morel, A non-local algorithm for image denoising, *IEEE Computer Society Conference on Computer Vision and Pattern Recognition,* CVPR, IEEE Computer Society, pp 60–65, (2005).
5. Y. Chen, T. Pock, Trainable Nonlinear Reaction Diffusion: A Flexible Framework for Fast and Effective Image Restoration, *IEEE Trans Pattern Anal Mach Intell* 39:1256–1272 (2017). doi: 10.1109/TPAMI.2016.2596743.
6. K. Dabov, A. Foi, K. Egiazarian, Image Denoising by Sparse 3-D Transform-Domain Collaborative Filtering, *IEEE Trans Image Process* 16:2080–2095 (2007).
7. I. Daubechies, The Wavelet Transform, Time-Frequency Localization and Signal Analysis, *IEEE Trans Inf Theory* 36:961–1005 (1990). doi: 10.1109/18.57199.
8. M.N. Do, M. Vetterli, The contourlet transform: An efficient directional multiresolution image representation, *IEEE Trans Image Process* 14:2091–2106 (2005). doi: 10.1109/TIP.2005.859376.
9. M. Elad, M. Aharon, Image Denoising Via Sparse and Redundant Representations Over Learned Dictionaries, *IEEE Trans Image Process* 15:3736–3745 (2006). doi: 10.1109/ICIG.2009.101.
10. K. Engan, S.O. Aase, J.H. Husoy, Method of Optimal Directions for frame design. ICASSP, *IEEE Int Conf Acoust Speech Signal Process - Proc* 5:2443–2446 (1999). doi: 10.1109/icassp.1999.760624.
11. L. Fang, W. Xianghai, Adaptive Total-Variation and Nonconvex Low-Rank Model for Image Denoising, *Int J Image Graph* (2023). doi: 10.1142/S0219467825500160.

12. S. Han, H. Mao, E. Gong, S. Tang, W.J. Dally, J. Pool, J. Tran, B. Catanzaro, S. Narang, E. Elsen, B. Research, P. Vajda, P. Facebook, DSD: Dense-Sparse-Dense Training for Deep Neural Networks (2017).
13. J.S. K. He, X. Zhang, S. Ren, Deep residual learning for image recognition, *IEEE Conference on Computer Vision and Pattern Recognition*. pp 770–778 (2016).
14. P.V. Lavanya, C.V. Narasimhulu, K.S. Prasad, Log-Gabor filter-based DT-CWT and Moth Flame-Lion optimization algorithm for image denoising with deep belief networks, *AIP Conf Proc* 2407:020028 (2021). doi: 10.1063/5.0074527.
15. M. Masood, T.Y. Al-Naffouri, Sparse reconstruction using distribution agnostic bayesian matching pursuit, *IEEE Trans Signal Process* 61:5298–5309 (2013). doi: 10.1109/TSP.2013.2278814.
16. S.S. Nisha, S.P. Raja, A. Kasthuri, Static Thresholded Pulse Coupled Neural Networks in Contourlet Domain - A New Framework for Medical Image Denoising, *Int J Image Graph* (2020). doi: 10.1142/S0219467820500254.
17. S.K. Panigrahi, S. Gupta, Joint bilateral filter for signal recovery from phase preserved curvelet coefficients for image denoising, *Int J Image Graph* (2021). doi: 10.1142/S0219467821500492.
18. A. Qayyum, A.S. Malik, M. Naufal, M. Saad, M. Mazher, F. Abdullah, T.A.R.B.T. Abdullah, Designing of overcomplete dictionaries based on DCT and DWT, *IEEE Student Symposium in Biomedical Engineering and Sciences*, pp 134–139 (2016).
19. Y. Quan, Y. Chen, Y. Shao, H. Teng, Y. Xu, H. Ji, Image denoising using complex-valued deep CNN. *Pattern Recognit* 111:107639 (2021). doi: 10.1016/J.PATCOG.2020.107639.
20. M. Scetbon, M. Elad, P. Milanfar, Deep K-SVD Denoising, *IEEE Trans Pattern Anal Mach Intell X* (2019).
21. J.L. Starck, E.J. Candès, D.L. Donoho, The curvelet transform for image denoising, *IEEE Trans Image Process* 11:670–684 (2002), doi: 10.1109/TIP.2002.1014998.
22. S.I. and C. Szegedy, Batch normalization: Accelerating deep network training by reducing internal covariate shift, *International Conference on Machine Learning*. pp 448–456 (2015).
23. A.K. Verma, B.S. Saini, T. Kaur, Image denoising using Alexander Fractional Hybrid Filter, *Int J Image Graph* (2018). doi: 10.1142/S0219467818500031.
24. D. Yang, J. Sun, BM3D-Net: A Convolutional Neural Network for Transform-Domain Collaborative Filtering, *IEEE Signal Process Lett* 25:55–59 (2018). doi: 10.1109/LSP.2017.2768660.
25. X. Yang, Y. Xu, Y. Quan, H. Ji, Image Denoising via Sequential Ensemble Learning, *IEEE Trans Image Process* 29 (2020).
26. K. Zhang, W. Zuo, Y. Chen, D. Meng, L. Zhang, Beyond a Gaussian denoiser: Residual learning of deep CNN for image denoising, *IEEE Trans Image Process* 26:3142–3155 (2017). doi: 10.1109/TIP.2017.2662206.
27. K. Zhang, W. Zuo, L. Zhang, FFDNet: Toward a fast and flexible solution for CNN-Based image denoising. *IEEE Trans Image Process* 27:4608–4622 (2018). doi: 10.1109/TIP.2018.2839891.
28. H. Zou, R. Lan, Y. Zhong, Z. Liu, X. Luo, EDCNN: A Novel Network for Image Denoising, *IEEE International Conference on Image Processing* (2019).

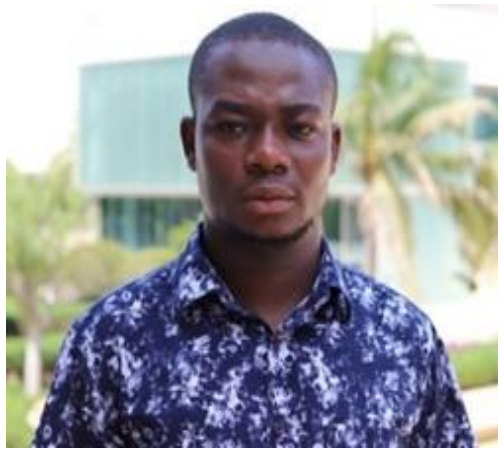

**Basit Alawode** received his First-Class BSc degree in Electrical/Electronic Engineering from the Obafemi Awolowo University (OAU), Ile-Ife, Nigeria, in 2013. He worked in several companies as an embedded systems designer and a software developer between 2013 and 2018. He obtained his fully-funded MSc degree in Electrical Engineering at the King Fahd University of Petroleum and Minerals (KUPM), Dhahran, Saudi Arabia in 2020. During this period, he performed several researches in the application of machine and deep learning (ML/DL) techniques to solve several challenges. He was a research intern at the King Abdullah University of Science and Technology, Thuwal, Saudi Arabia in 2021. His research interests include the application of ML/DL to power, control, embedded, signal, and image processing systems. He is currently pursuing his Ph.D. in engineering at Khalifa University, Abu Dhabi, United Arab Emirates where he is researching on underwater computer vision systems and applications.

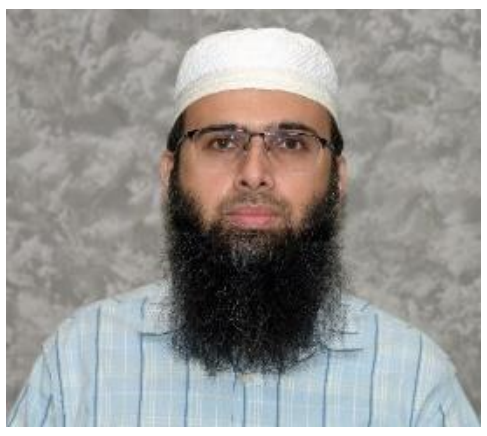

**Mudassir Masood** received the B.E. degree in computer systems from the NED University of Engineering and Technology, Karachi, Pakistan, in 2001, the M.S. degree in electrical engineering from the King Fahd University of Petroleum and Minerals (KFUPM), Dhahran, Saudi Arabia, in 2005, and the Ph.D. degree in electrical engineering from the King Abdullah University of Science and Technology (KAUST), Thuwal, Saudi Arabia, in 2015. He was a Post-Doctoral Researcher with Texas AM University at Qatar, Doha, Qatar, from 2015 to 2016. He is currently an Assistant Professor with the Electrical Engineering Department, KFUPM and is also affiliated with the Center of Communication Systems and Sensing. His research interests lie in the areas of machine learning, image processing, compressive sensing, massive MIMO, physical layer security, and related applications. He was a recipient of a best student poster award at ICCSPA2013, and the KAUST Fellowship and the Academic Excellence Award 2010.